# Phase Transition in a Memristive Suspended MoS$_2$ Monolayer Probed by Opto- and Electro-Mechanics.

Julien Chaste[1*], Imen Hnid[1], Lama Khalil[1], Chen Si[2], Alan Durnez[1], Xavier Lafosse[1], Meng-Qiang Zhao[3], A.T. Charlie Johnson[3], Shengbai Zhang[4], Junhyeok Bang[5**], Abdelkarim Ouerghi[1]

[1]Université Paris-Saclay, CNRS, Centre de Nanosciences et de Nanotechnologies, 91120, Palaiseau, France.
[2] School of Materials Science and Engineering, Beihang University, Beijing 100191, China
[3] Department of Physics and Astronomy, University of Pennsylvania, 209S 33rd Street, Philadelphia, Pennsylvania 19104 6396, United States
[4] Department of Physics, Applied Physics, & Astronomy, Rensselaer Polytechnic Institute, Troy, New York 12180, USA
[5] Department of Physics, Chungbuk National University, Cheongju 28644, Republic of Korea

* Corresponding author: julien.chaste@universite-paris-saclay.fr
* Corresponding for theory: jbang@cbnu.ac.kr

**Abstract**
Semiconducting monolayer of 2D material are able to concatenate multiple interesting properties into a single component. Here, by combining opto-mechanical and electronic measurements, we demonstrate the presence of a partial 2H-1T' phase transition in a suspended 2D monolayer membrane of MoS$_2$. Electronic transport shows unexpected memristive properties in the MoS$_2$ membrane, in the absence of any external dopants. A strong mechanical softening of the membrane is measured concurrently and may only be related to the phase 2H-1T' phase transition which imposes a 3% directional elongation of the topological 1T' phase with respect to the semiconducting 2H. We note that only a few percent 2H- 1T'phase switching is sufficient to observe measurable memristive effects. Our experimental results combined with First-principles total energy calculations indicate that sulfur vacancy diffusion playsa key role in the initial nucleation of the phase transition. Our study clearly shows that nanomechanics represents an ultrasensitive technique to probe the crystal phase transition in 2D materials or thin membranes. Finally, a better control of the microscopic mechanisms responsible for the observed memristive effect in MoS$_2$is important for the implementation of future devices.

Semiconducting two-dimensional transition metal dichalcogenides (2D TMDs) are promising candidates for optoelectronic devices and electronic transistors.[1,2] Their 2D nature strongly affects the value of the excitonic binding energy,[3] spin-orbit coupling,[4] and energy band gap.[5] These substantial changes of the electronic characteristics turn 2D $MoS_2$ in a versatile material which combines multiple interesting properties. For example, 2D $MoS_2$ has been recently investigated both for memristive devices.[6,7] and energy storage applications.[8–11]

A memristor is a circuit element[12] combining both memory and transduction[13]. Most memristor relies on dopant migration to alter their electronic properties, *e.g.* $Li^+$ ion migration in multilayer $MoS_2$ or oxide vacancies in $TiO_2$[13]. For multilayer $MoS_2$, a strong doping due to the intercalation of 0.4 Li atom per unit cell can induce a crystal phase transition between the 1T' and 2H crystalline phases of the 2D TMD material,[14] and thereby creating interesting memristive phenomena. In the absence of strong dopants, other memristive effects have been observed in lateral[6] and vertical $MoS_2$-based transistors.[9,15] and linked to the migration of grain boundaries and sulfur vacancies ($S_V$). However it is unclear if those models can apply to the simple case of a single monocrystalline $MoS_2$ flake of monolayer thickness, for which grain boundaries and intercalation phenomena are absent.

Standard characterization methods, such as Raman spectroscopy,[16] photoemission spectroscopy or transmission electron microscopy (TEM),[17,8] have been employed to study both 2H and 1T' phases in TMDs. However, all these methods present some drawbacks, and they are difficult to combine with *in situ* electrical measurements of the memristive effect. For example, TEM unambiguously discriminates the 2H and 1T' phases but it is restricted to a small area and can affect the dynamics of crystal phase transition depending on the beam excitation. Similarly, optical methods based on lasers can also induced such artefacts with power above $100\mu W$[18]. Consequently, there is a need for atypical characterization methods, which can observe and probe both the phase transition and the vacancy migration in TMDs in a non-disturbing manner. Since flexural vibrations are very sensitive to force, crystal distortion, charge or atoms absorption,[19] an appropriate way to explore in detail the behaviors of 2D TMDs is nanomechanics. The emergence of suspended materials with a single-atom thickness[20] brought interesting properties for nanomechanical resonators including very low mass and stiffness constants,[21] high resistance to elongation, variable frequency, and strong mechanical non-linearities.[22] Nanomechanical resonators made of ultrathin suspended monolayers of TMDs [23–26] may be less studied than their graphene counterparts, but they are also more suitable for technological applications.

In this work, we probe by nanomechanics the memristive properties of a suspended monocrystalline $MoS_2$ flake of monolayer thickness which has a high density of $S_V$ defects (10-18%, determined in a previous report)[27]. The device exhibits persistent photocurrents (PPCs) related to $S_V$ dopants and their slow diffusion on few micrometers. This scenario is totally different when probed at the nanometer scale and our findings unveil that S vacancies have a faster and high in-plane diffusivity, and that they can promote the nucleation of the 1T' phase. With a simple voltage and optical excitation, we are able to control the $S_V$ drift over nanometric scales, it is enough to imply a phase transition and a resulting memristive effect. These results are confirmed by first-principles total energy calculations, which show that the optical excitation can significantly enhance the $S_V$ diffusion and accelerate the phase transition from the 2H phase to the 1T' one. This phase transition was only detected for a very small part of the material with a strong softening of the membrane.

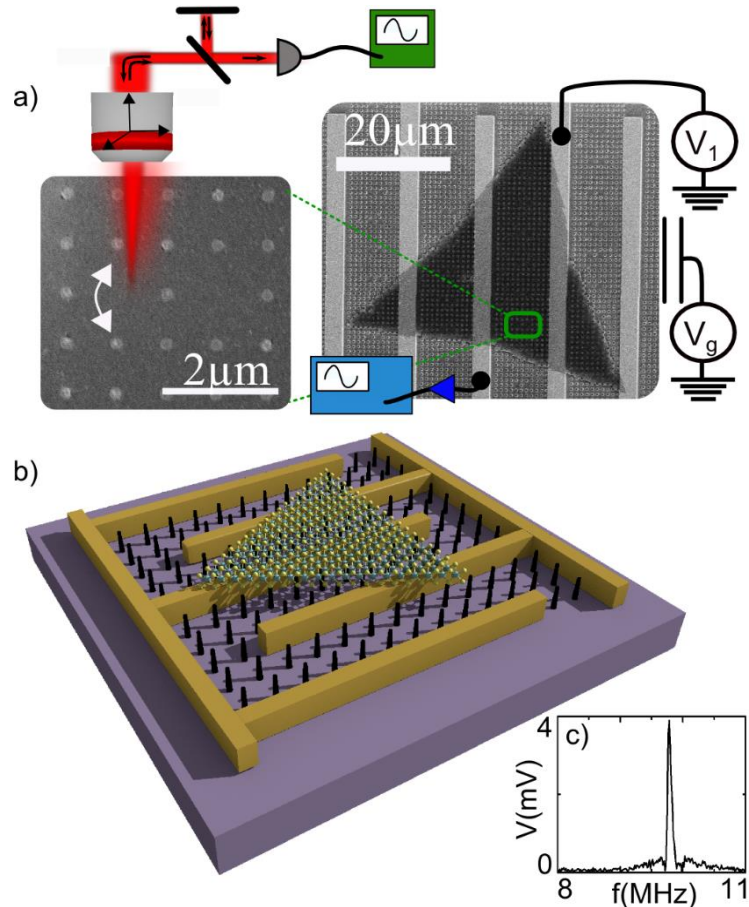

**Figure 1 Optomechanical and electrical set-up for the suspended MoS₂ membrane.** a) Electronic image of a monolayer MoS$_2$ suspended between interdigitated contacts and an array of SiO$_2$ nanopillars. It resulted in a periodic series of local mechanical oscillators. The vibration was excited with an electrostatic back-gate and measured with a reflective laser (633nm,) using a Michelson interferometer. b) Schematic of the sample showing the MoS$_2$ flake suspended above the pillar array. c) Typical mechanical resonance measurement.

**Results and discussions**

Figure 1a presents the geometry of our monocrystalline single-layer MoS$_2$ flake, grown by chemical vapor deposition (CVD),[28] suspended on an array of SiO$_2$ pillars and interdigitated electrical contacts. The specimen was embedded in a nano-opto-electro-mechanical system which uses electrocapacitive excitations to induce periodic mechanical vibrations in the MoS$_2$ monolayer, finally detected by optical interferometry (Figure 1b). The arrangement of the metallic electrodes and SiO$_2$ substrate provides the electrostatic gating of the MoS$_2$ flake. We measure the vertical motion of the membrane with a reflected laser signal and its vibrations by means of a Michelson interferometer, as shown in Figure 1a. We observe mechanical oscillations centered at $f_0 = 1/2\pi \sqrt{k/m}$ at ~ 10 MHz with k and m the resonator spring constant and the effective mass of the resonator, respectively. The resonance quality factor (Q) is 320 at room temperature, in agreement with our previous report.[29]

The electrical transport properties of the suspended MoS$_2$ film were then investigated for different fluence of a red laser source (Figure 2a). We notice that the conductance is enhanced with increasing laser power due to a photodoping effect related to the presence of in-gap states in MoS$_2$.[30] Without any light, a robust persistent photocurrent remains present for few hours in our sample.

Beside this persistent photocurrent, the other feature is a clear hysteretic loop in I–V when $V_{ds}$ is swept back and forth from -7V to 7V, which reflects a nonlinear memristive behavior of our system. The current sweeps between two states 0 and 1 at a threshold voltage $V_H$ and at a current step $\Delta I_H$. Note that our device is electrically bipolar (see see Supplementary Information Figure S1).

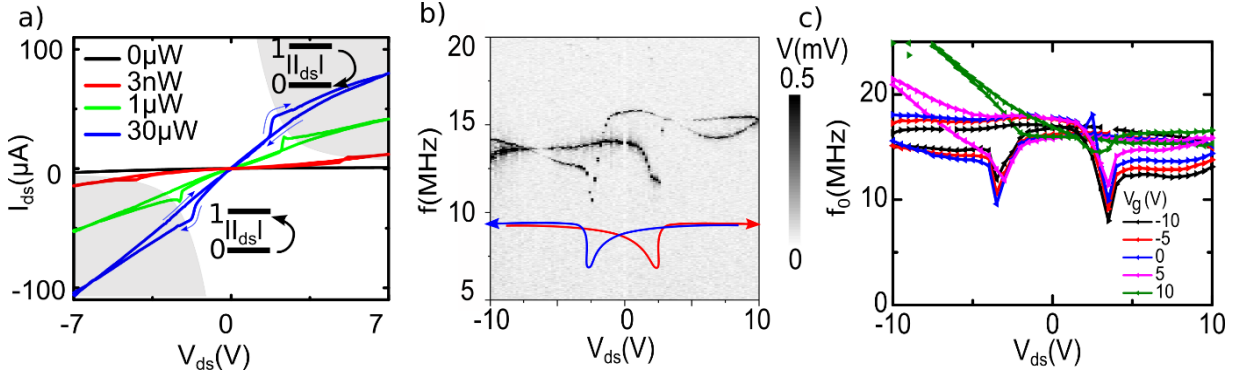

**Figure 2 Memristive behavior of the MoS$_2$ for both current and mechanical resonance as a function of $V_{ds}$.** a) The $I_{ds}(V_{ds})$ for both sweep directions and for different laser powers presents a clear hysteresis and memristive property. In the white (grey) region, the current tended to increase (decrease) along the time sweep, at fix $V_{ds}$, and the memristor current went from state 0 to 1 (1 to 0). b) The laser reflection displayed an optomechanical response to an electrical excitation at frequency f and was traced as a function of $V_{ds}$ for both sweep directions. We observed a variation of the resonant frequency $f_0$ with a hysteresis along $V_{ds}$ and a strong softening of the mechanical mode. The frequency was reduced by half in this case at the same polarization as the memristive effect. Red and blue lines have been drawn for clarity. c) The mechanical frequency of the same resonator as a function of $V_{ds}$ for different back-gate voltage.

Further insights into these memristive effects, can be obtained by performing nanomechanical measurements.. Figures 2b and 2c reveal the mechanical resonance as a function of $V_{ds}$. As one can see, the mechanical mode in the hysteretic loop displays a strong softening, attributed to the memristive behavior of MoS$_2$ at $\pm V_H$. In addition, the resonant frequency at $V_{ds} = 0V$ is reduced by more than two times with respect to the minimum frequency, $f_{min}$ (Figure 2c); the ratio $f_{0,V_{ds} = 0V}/f_{min}$ was determined to be equal to 2.7 (see Supplementary Information Figure S2a). In contrast to the electronic, the mechanical softening seems here unipolar (see Supplementary Information Figure S2c). Since we have previously measured the strain in our MoS$_2$ membrane equal to 0.05%, [29] the motion can be described within the stress limit, and hence the frequency is of the form $f_0 = 1/2\pi . \sqrt{2.4048^2.T/(R^2\rho)}$ where $\rho=3073$ kg.m$^{-2}$ refers to the 2D mass density of MoS$_2$, R is the radius of the membrane, and T is the total stress (in N/m) applied at the end of the membrane. Only one explanation is relevant here; a 2H-1T' phase transition of only 1% is capable of inducing this strong softening and is sufficient to reduce the total deformation of the membrane to zero. Also, this transition, 2H-1T' naturally yields a shift of conductance, from a high gap to a small gap semiconductor, as obtained in the electrical measurements. In contrast, S$_V$ centers are assumed to stress locally on the material[31], which is in contradiction with our observations. A downshift of the frequency related to the mass is excluded, see S4. In Figure 2c, besides, the lack in the gate voltage dependence at $V_H$ strongly suggests that the memristive effect is not a capacitive coupling (see Supplementary Information Figure S4).

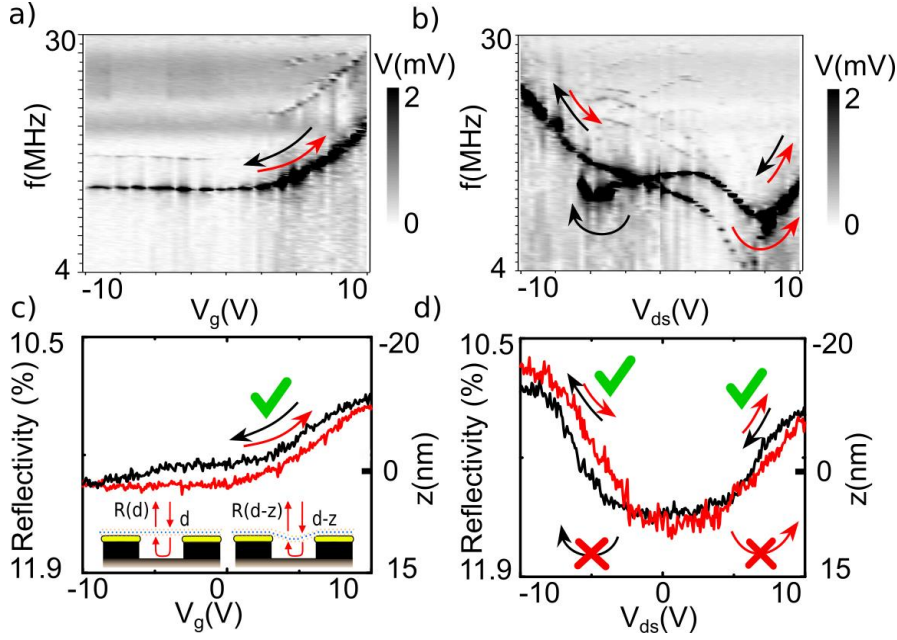

**Figure 3 Strain engineering in MoS$_2$.** We measured the optomechanical response for $V_g$ (a) and $V_{ds}$ (b) and the two sweeping directions. The optical reflectance (633nm) in percent of the input power as a function of $V_g$ (c)($V_{ds}$=-5V) and $V_{ds}$ (d) ($V_g$=4V) for both sweeping directions (the black curve corresponds to a decreasing potential and the red curve to an increasing one). Inset: schematic of the capacitive electrostatic force applied to the 2D membrane, the reflectivity changes with the distance d when a potential is applied. Both measurements displayed similar dependencies but the hysteresis in $V_{ds}$, corresponding to the memristive effect, was not present in the reflectance. The same optomechanical measurements were performed before and after the reflectance measurements to confirm that it was not an artifact or a deviation of the xyz position.

To confirm that MoS$_2$ exhibits a phase transition, which create a stress in the membrane plane, we described our resonator as a circular membrane with clamped sides. We considered the total stress to be $T(z)=T_0+T_D(z)$, where $T_0$ is the built-in stress and $T_D$ is the stress due to any deflection induced by electrostatic forces. We defined z as the deflection at the center of our circular membrane (r = 0). The frequency for the fundamental mode, given by the Euler-Bernoulli description, is defined by (see Supplementary Information Figure S3):

$$f_0=\frac{1}{2\pi}\sqrt{2.405^4\frac{Et^3}{12(1-\nu^2)}\frac{1}{R^4\rho}+2.405^2\frac{T_0}{R^2\rho}+2.405^2\frac{12}{3}\frac{Et}{(1-\nu^2)}\frac{z(V'_g)^2}{R^4\rho}-1.23\frac{\epsilon_0 V'^2_g}{d^3\rho}} \quad (1)$$

We define $\nu$ as the Poisson ratio, E the Young modulus of the material and $\epsilon_0$ its permittivity. $V'_g$ is a renormalized gate potential, $V'_g \sim V_g - V_{ds}/2 - \mu/e$, where $\mu$ is the chemical potential of the MoS$_2$ and e the elementary charge. It is noteworthy that the first term in $Et^3$, *i.e.* the material bending is negligible (~200 kHz in our conditions). The second one in $T_0$ corresponds to the in plane built-in stress of the membrane, and the last two involve a capacitive pull-in (V'g) with the out of plane deflection (see Supplementary Information Figure S4).

In order to differentiate the in-plane from out of plane movements, we have measured the z variation along the sample by optical reflectometry. The optical cavity formed by the suspended MoS$_2$ and the silica substrate is of bad quality, but we still manage to measure a deflection along z with approximately 0.04% per nm change of the reflectivity signal (see Supplementary Information Figure S5). In Figure 3, we acquire simultaneously the mechanical frequency (Figure 3.a-b) and the laser reflectivity (Figure 3.c-d), as a function of $V_g$ and $V_{ds}$, in the same conditions. We first observe a vertical deflection of 10-20 nm for high $V_{ds}$ and $V_g$ in both sweep directions (Figure 3.c-d). A membrane hardening corresponding to the third term in equation 1.

Surprinsigly, , while the memristive softened mode is well visible in the mechanical resonance (Figure 3b), no corresponding signature appears in the vertical (Figure 3d). In the regime dominated by an in-plane intrinsic strain at $V_{ds} \sim V_g = 0$, the static deflection is $z \approx 0$ nm. The modification of this in-plane strain does not affect z and the reflectance, but only the dynamical oscillating deflection $z_{AC}(t)$ resulting from the forced vibration. Therefore, we can conclude that the memristive visible in the mechanical resonnance is related to an in plane deformation of the membrane, as in the case of the 2H-1T' phase transition of $MoS_2$.

Previous investigations have shown that the migration, diffusion or/and clustering of $S_V$ defects are at the heart of memristive behaviors,[6,15,32] photodoping[30] and the 2H-1T' phase transition[8,31] in $MoS_2$. The presence of such abundant and stable $S_V$ vacancies can create localized in-gap states and thus a *n*-type doping in the layer,[27] in which the injection is of 0.04e per vacancy.[31] Obviously, for our suspended membrane, persistent photocurrents (PPCs) are not related to the substrate, [33,34] but to intrinsic defects, *i.e.* $S_V$ centers. Therefore the measurement of those PPCs can help to understand the dynamics of $S_V$ acancies over a long time scales (around 500s) [30,35] and migration length (typically 1µm). We have probed the PPCs for different voltages $V_{ds}$ in Figure 4j (see also Supplementary Information Figure S7). The normalized $I_{ds}$ curves were fitted with a sum of exponential functions $I(t)= I_1 e^{-t/\tau_1}+I_2 e^{-t/\tau_2}+\Delta I e^{-t/\tau_3}$, where $\tau_1$ and $\tau_2$ are the characteristic decay times related to two different types of charge traps in $MoS_2$ [33,34] ( $\tau_1=269 \pm 48$ s and $\tau_2$ 8815 $\pm$ 690 s). $\tau_3$, on the other hand, is much faster and relates to the memristive effect ($\tau_3 \sim 20$ s) and is only included if $V_{ds}>V_H$.

In Figure 4i, we observe a similar current $\Delta I$ when the light is ON but greatly hidden by the photocurrent $I_{photo}$. We found a solution to highlight this effect; we explore the long-time behavior of vibrations. Since we attribute this effect and PPC to vacancies which tends to strain mechanically the $MoS_2$, it seems quite a natural method.

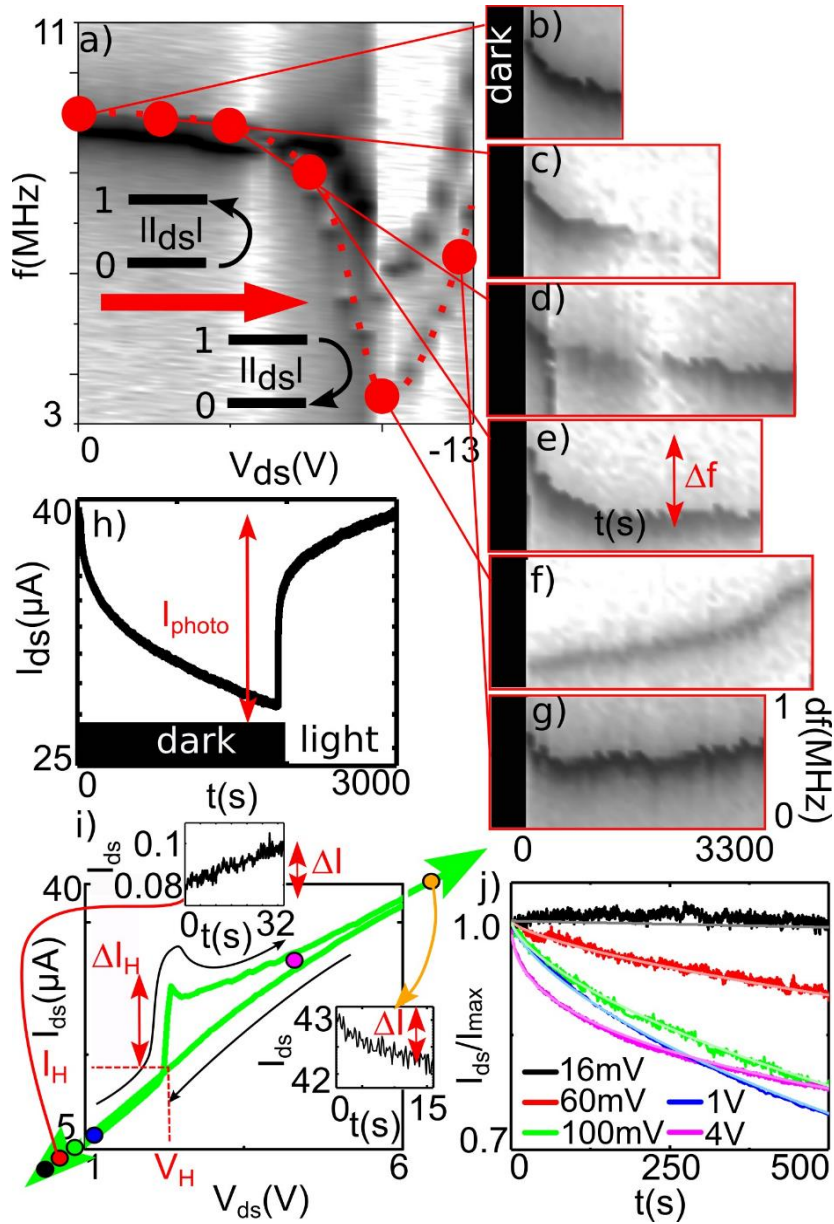

**Figure 4 Correlation between the current and the mechanical frequency.** In a, we traced the optomechanical response of our system for the fundamental mode of two adjacent resonators, with $V_{ds}$ increasing (the reverse sweeping was in SI). We observed a softening of the mechanical resonance mode from 9.2 to 3.4 MHz. From b to g, the temporal evolution of the mechanical response was measured at different values of $V_{ds}$. The resonant mode slowly decreased for about 400-500s, by $\Delta f$, for $V_{ds}<V_H$ and tended to increase for the higher values $V_{ds}>V_H$. This dynamic was strongly correlated with the current properties of our sample. h)The slow time evolution of the current $I_{ds}$ which decreased by $I_{photo}$ in the dark and increased under the laser light, with $V_{ds}> V_H$ i) A zoom of the $I_{ds}(V_{ds})$ curve of Figure 2 around the memristive charge accumulation represented by a current step $\Delta I_H$ at a threshold position ($V_H$, $I_H$). In the two insets, for both regimes, at a fixed $V_{ds}$ value above and below $V_H$, the evolution during the first tens of seconds of the current when the light was already ON. Below the $V_H$ threshold, $I_{ds}$ increased by $\Delta I$ and above $V_H$, $I_{ds}$ decreased in strong analogy with the -$\Delta f$ dependence described in (a-e) and (f-g). After almost 100s, the current tended to increase again in both cases due to $I_{photo}$, in accordance with (h). j) Time relaxation of the normalized photocurrent at different $V_{ds}$ with respective exponential decay fit

In Figures 4a-g, as for the PPC, we have identified long-lasting dynamics for $S_V$ defects *via* nanomechanical measurements. They are also correlated to memristive effect. After keeping our device in the darkfor 30 minutes, we switched ON the laser onto our sample and immediately measured the vibration and its temporal evolution at different values of $V_{ds}$. Surprisingly, at low voltages, the mechanical frequency downshifted by $\Delta f = 0.3\text{-}0.8$ MHz over a long period of time. This is similar to the PPC with typical timescale close to $\tau \sim 400\text{-}500$s and a modification at $V_H$ and above; the evolution of $\Delta f$ at long delays corresponds this time to a strong hardening of MoS$_2$. We point out the strong analogy between $\Delta f$ and $\Delta I$, and their relation to the memristive effect, $\Delta f$ being without significant parasitic contributions. This slow effect seems intrinsically different from the fast memristive effect observed in I–V in Figure 2 and to the phase transition of MoS$_2$ but related to it. We believe this antagonism lies in their common point; the $S_V$ vacancies. Indeed, for the same $S_V$ diffusion in $L^2/\tau$, we can have completely different dynamic $\tau$ if we consider different length L as the sample length in micrometer and a local, nanometric, effect. PPC and memristive effect are simply two different consequences of $S_V$ diffusion over different scales.

All the above-mentioned results link the memristive effect, induced by a phase transition, to the diffusion of the $S_V$ defects over a nanometric lenght. From litterature, the relation between $S_V$ and the 2H-1T' phase transition first lies in the capacity of $S_V$ centers to agglomerate in lines, referred as $S_v$ clusters, or in an intermediate S-poor α-phase[8,31]. Additionally, dynamics of the defective centers can also play a role, as a line of vacancies can release the strong repulsion at the 2H-1T' interface (from a 2H phase to a 1T phase, S atoms are sliding together on one side of the 2D material). The dynamics of $S_V$ clustering and subsequent local 1T' phase transition are examined here using first-principles calculations. Figures 5a and 5b represent the $S_V$ diffusion pathway and the corresponding barrier. We find that in the electronic ground state, the diffusion barrier is quite high (2.3 eV), and Sv is thus an immobile species. When the electron is excited by laser irradiation, the diffusion barrier is reduced to 1.04 eV and the Sv diffusion is facilitated.

Such a Sv diffusion can lead to a vacancy clustering.[36,37] Using a rate equation, one can quantitatively understand the dynamics of Sv clustering. We consider three processes: (1) the diffusion of a single S vacancy (1Sv), (2) the vacancy clustering [1Sv + nSv → (n+1)Sv], and (3) the dissociation of a vacancy cluster [nSv → 1Sv + (n-1)Sv]. Here, nSv denotes the n S vacancy cluster. The time variation of the nSv concentration, denoted as [nSv], is calculated using the following coupled equations:

$$\frac{d[1Sv]}{dt} = -k_O[1Sv][1Sv] - \sum_{n=1} k_o[1Sv][nSv] + k_d(2Sv)[2Sv] + \sum_{n=2} k_d(nSv)[nSv]$$

$$\frac{d[nSv]}{dt} = k_O[1Sv][(n-1)Sv] - k_o[1Sv][nSv] + k_d\big((n+1)Sv\big)[(n+1)Sv] - k_d(nSv)[nSv], for n \geq 2$$

Because the ordering is determined by the diffusion 1Sv, the clustering rate $k_o$ can be calculated by

$$k_o = 2\pi a^2 f exp\left[\frac{-E_{diff}}{k_B T}\right]/ln\left[\frac{1}{a\sqrt{\pi[1Sv]}}\right].$$

Here, $E_{diff}$ is the diffusion energy barrier of 1Sv, $a$ is the MoS$_2$ lattice constant, and $f$ is the vibrational frequency.[38] The dissociation rate is $k_d = f exp[-E_{dis}(nSv)/k_B T]$, where $E_{dis}(nSv)$ is the dissociation energy barrier, given by the summation of $E_{diff}$ and the binding energy between 1Sv and (n-1)Sv.[36,39] We use the calculated binding energies for the di-, tri-, and tetravacancy, that is, 0.08, 0.19, and 0.25 eV,[11] respectively, but set to 0.30 eV for the binding energies of larger vacancy clusters.

As an initial condition, we set [1Sv] = 2.0x10$^{13}$ cm$^{-2}$, which corresponds to 1% 1Sv vacancies, and [nSv] = 0 for n ≥ 2. Figure 5c shows that, in the electronic ground state, no vacancy clustering is found until 0.1 s, because of the low diffusivity of 1Sv. On the other hand, Figure 5e shows that, in the electronic excited state, the enhanced diffusivity of 1Sv leads to large vacancy clustering. Such a vacancy cluster nSv, is associated to a local 1T' phase within an area of $A = \sqrt{3}/4 \cdot (na)^2$ .[37] (see Figure 5c). Thus, the ratio of the total area of the 1T' phase $A_{1T'}$ to the area of the sample $A_{sample}$ is given by

$$A_{1T'}/A_{sample} = \frac{\sqrt{3}}{4} \sum_{n=2} [nSv](na)^2.$$

Using these reaction rates, we show in Figure 5f that approximately 5% of the 2H MoS$_2$ area can transition to the 1T' phase within 100 ms.

Under simulation, we have established, under laser illumination, a strong reduction of the barrier energy for diffusion of vacancies, from 2,3 eV to 1,04 eV and the enhancement of S$_V$ clustering when the density is only 1% in the material. It is in accordance with experimental results showing E is reduced even further, to 0.5 to 0.8 eV, when the S$_V$ concentration is higher[35,40,41] and closer to the 18% in our MoS$_2$. The 1T' phase was observed after the incorporation of S vacancies in ref.[31]

We also confirm the trace of few percent of the 1T' phase in our sample through Raman and photoemission spectroscopy (see Supplementary Information Figure S9). During the voltage sweeping, the S vacancies tend to conglomerate in clusters or lines rather than be spread out when subjected to an electronic potential[8,11] and form, in a reversible way, the boundary precursor of the 1T phase. Along the measurement, we modulated experimentally by only 1% the density of 1T'phase but it is enough to induce the memristive effect and the mechanical softening. We show that the control over S$_V$ diffusion is a substitute to usual doping method inducing a phase transition and fully tunable in real devices.

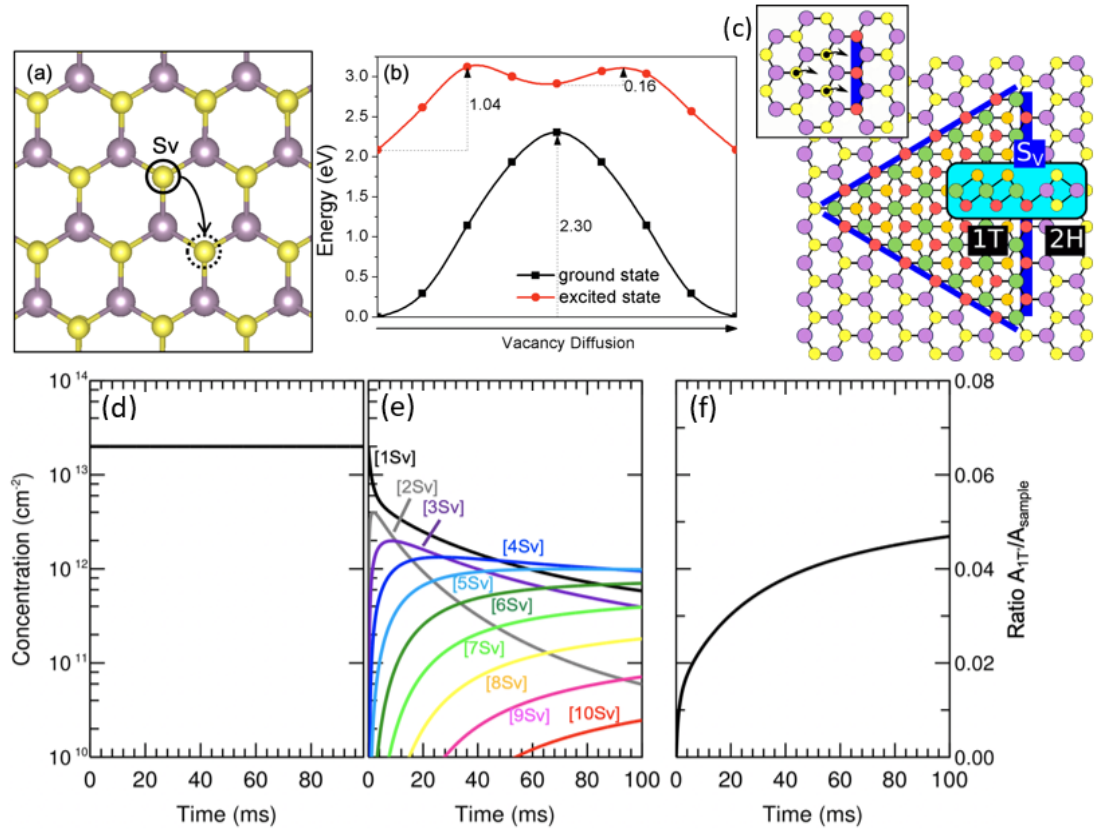

**Figure 5** (a) Pathway schematics and (b) energy variation of the 1Sv diffusion. In (b), black and red lines stand for the electronic ground and excited states, respectively. c) Schematic for the formation of a 1T phase triangular shape into the natural 2H phase, due to mainly sulfur displacement, and allowed by the presence, at minima, of boundaries lines of Sulfur vacancies. Time evolution of the concentration of the n vacancy cluster [nSv] in (d) ground and (e) excited states. (f) The change ratio of the total area of the 1T' phase $A_{1T'}$ to the area of the sample $A_{sample}$ by the vacancy clustering in the excited state, *i.e.*, in (e).

To finish the discussion, we comment here about the slow time relaxation of $S_V$ centers (t = 500s) for PPC and the voltage dependence. We naively defined the self-diffusion D of $S_V$ centers along the $MoS_2$ at equilibrium[42] by: $D = a^2. k_B T/h . exp(-E_{diff}/k_B T) 10^{-15} - 10^{-20} m^2. s^{-1}$. h is the Planck constant and $E_{diff}$ is 0.5 to 0.8 eV.[35,40,41] Another way to define D is $D \sim L^2/\tau \sim 10^{-15} m^2. s^{-1}$ with the typical length of our samples, L = 1 µm. It is in total agreement with reported values of D = 3,8.10$^{-22}$ m$^2$.s$^{-1}$ (D = 5.10$^{-19}$ m$^2$.s$^{-1}$) under electron irradiation.[35,40] PPC and phase transition are related to $S_V$ diffusion, the difference is the locality of the latter with length around ~ 1 nm; the dynamic is therefore 6 order of magnitude faster for the transition and the memristive effect. When an electrical potential is applied, the Nernst-Einstein law define an equilibrium where the $S_V$ current compensate the diffusion. This relation defines an $S_V$ drift speed $v = De/k_B T . V_{ds}/\Delta L$ of 10 nm/s at $V_{ds}$= $V_H$~2.5V, which is favorable for a quick and local clustering of vacancies expected in the phase transition. Here, $\Delta L \approx 10 \mu m$ is the electrode distance and e is the elementary charge.

**Conclusion**

To conclude, we have demonstrated the relation between memristive effects in a monocrystalline $MoS_2$ device of monolayer thickness and an intrinsic crystal phase 2H-1T' transition using a simple yet effective optoelectromechanical technique. As mechanical vibrations are extremely sensitive to strain (ultimately $\Delta e \sim 10^{-10}$ with $\Delta f/f 10^{-6}$),[43] our method is non-disturbing, local and compatible with other *in situ* electronic and optoelectronic measurements.. The Sv diffusion was found to play a key role in the phase transition, which was evidenced *via* the slow dynamics of both PPCs and nanomechanical vibrations. Additionally, theoretical calculations prove that a high reduction in the energy barrier of $S_V$ diffusion is important to yield the phase transition, due to $S_V$ clustering under laser irradiation. Suspended single layer $MoS_2$ with a high $S_V$ concentration is therefore a promising and low-cost system for future applications, in particular in the electrochemical hydrogen production. Controlling the diffusion of the Sv defects appears to be the main tool to exploit all properties of $MoS_2$, as well as other 2D TMDs.

**Experimental section;**
**Fabrication method**
For the sample fabrication, we embedded a large interdigitated gold metallic contact, 50nm, in a $Si/SiO_2$ substrate by dry etching and e-beam lithography in order for the top of the contact to be at the same level as the substrate surface. The square array of pillars (pitch 1 micron) was created by dry etching of the $SiO_2$, with an evaporated hard Ni mask (70nm). A PMMA resist was spin-coated onto the as-grown $MoS_2$ flakes resting on the $SiO_2$ substrate using during CVD growth.[44,45] Large monocrystalline flake of monolayer $MoS_2$, together with the resist layer, were detached from the growth substrate by wet etching in a KOH solution and rinsed in deionized water. Still in the water, the PMMA layer containing the $MoS_2$ flakes layer was deposited on the array of pillars between the electrical contacts. The resist was then removed with acetone and the sample was finally dried with a critical point dryer.

**Experiment**
Electrostatic gating was provided by the doped Si substrate. For the measurement, we used a capacitive excitation and an optomechanical detection according to ref.[25] We measured the reflected signal of a laser (633nm, ~10μW) with both a Michelson interferometer and *in situ* Fabry-Perot cavities formed by the space between the substrate and the $MoS_2$. All the measurements were performed at ambient temperature under vacuum. The vibration responsivity of the optical cavity $MoS_2$-substrate had been optimized for enhanced reflectance (see supplementary information).

**Calculation method:**
The first principle calculations were performed using density functional theory[46] with the projected augmented wave formalism [47] as implemented in the Vienna *ab initio* simulation package. The exchange and correlation effects were described by the Perdew−Burke−Ernzerhof version of the generalized gradient approximation. [48] An energy cutoff of 400 eV was set for the plane wave expansion. We used a 5×5 supercell of monolayer $MoS_2$ with a single vacancy to simulate the vacancy diffusion in $MoS_2$. A 15 Å thick vacuum layer was included to avoid the interaction between the periodic slabs. To mimic the electronic excitation, we performed occupation-constrained DFT calculations [49,50] in which we removed electrons from the highest occupied band and placed them at the lowest unoccupied band. In addition, we assume the elevated temperature $T = 300$ ºC by the laser irradiation (see supplementary information).

**Acknowledgments:** We thanks Fabrice Oehler for his comments. Experimental work was supported by French grants ANR ANETHUM (ANR-19-CE24-0021) and ANR Deus-nano (ANR-19-CE42-0005), by the European Union's Horizon 2020 research and innovation program under grant agreement No 732894 (FET Proactive HOT) and by the French Renatech network. M.Q.Z. and A.T.C.J. acknowledges support from the Chinese/U.S. National Science Foundation EAGER 1838412 and MRSEC DMR-1720530. First-principles calculation was supported by National Natural Science Foundation of China (11874019) (C.S), Chinese/U.S. National Research Council of Science & Technology (No. CAP-18-05-KAERI) (J.B.), and U.S. Department of Energy (DOE) under Grant No. DESC0002623 (S.B.Z.).

**Supplementary information:** Additional information is available online. The Supplementary Information proposes additional measurements on memristive effect and PPC with some models. It also details a description of the movement equation of the resonator as well as a more complete discussion on the origin of the observed mechanical softening.

**Author contribution:** J.C. and A.O. conceived and designed the experiments, M.Q.Z, A.T.C.J., A.D., X.L. and J.C. designed and fabricated the devices, I.H., L.K. and J.C. performed the experiments, Q.Z., S.Z., J.B. developed the theory, and J.B. J.C., L.K. A.O. co-wrote the paper. All the authors discussed the results and commented on the manuscript.

**Supplementary information for:**

**Phase Transition in a Memristive Suspended MoS$_2$ Monolayer Probed by Opto- and Electro-Mechanics.**

Julien Chaste[1*], Imen Hnid[1], Chen Si[2], Lama Khalil[1], Alan Durnez[1], Xavier Lafosse[1], Meng-Qiang Zhao[3], A.T. Charlie Johnson[3], Shengbai Zhang[4], Junhyeok Bang[5**], Abdelkarim Ouerghi[1]

[1] Université Paris-Saclay, CNRS, Centre de Nanosciences et de Nanotechnologies, 91120, Palaiseau, France.
[2] School of Materials Science and Engineering, Beihang University, Beijing 100191, China
[3] Department of Physics and Astronomy, University of Pennsylvania, 209S 33rd Street, Philadelphia, Pennsylvania 19104 6396, United States
[4] Department of Physics, Applied Physics, & Astronomy, Rensselaer Polytechnic Institute, Troy, New York 12180, USA
[5] Department of Physics, Chungbuk National University, Cheongju 28644, Republic of Korea

* Corresponding author: julien.chaste@universite-paris-saclay.fr
* Corresponding for theory: jbang@cbnu.ac.kr

**Table of contents**

**S0: Sample images**

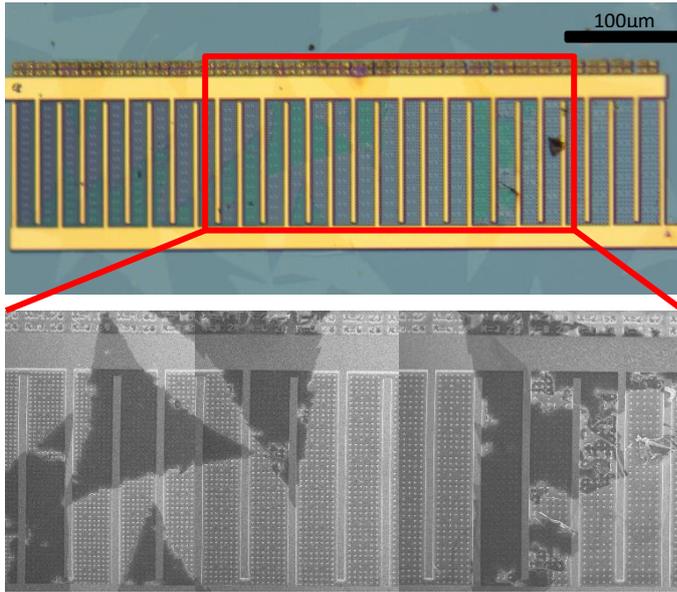

**Figure S0**; Optical image and E beam image of a typical sample used in this study, most of the $MoS_2$ flake are suspended over $SiO_2$ pillars and gold contacts

The $MoS_2$ sample presented in Figure S0 corresponds to most of the measurements achieved in the paper. One interest of this configuration, with interdigitated contacts is the possibility to optically illuminate one specific area of the sample and subsequently to electrically "activate" this same area due to the strong photocurrent in the sample.

**Convex edges and sulfur vacancies**
Another indirect consequence or signature of a high density of defect is the peculiar form of the $MoS_2$ triangle with convex edges. $MoS_2$ present in Figure S0 has this specific shape. Similar images on our $MoS_2$ with such type of edge are also observed in the SOI of the reference [1] and is well describe in reference. [2]

**S1: Electrical hysteresis and memristivity**
We measure the memristive effect in our devices and define the two states 0 and 1, they represent the maximum (minimum) current achievable for each $V_{ds}$ and P. The non-linearity of the $I_{ds}(V_{ds})$ curve is not as strong as previous memristive devices[3], even for $MoS_2$[4] but is significant. We present, in the Figure S2a, the two current states 0 and 1 and the protocol to pass from state 0 to state 1 and vice versa. We define two inflexion points P- and P+ (at -5.8V and +5.8V respectively for the sample in Figure S2a), two erasing area for $V_{ds}$ > P+ and $V_{ds}$<P- and a writing area P-<$V_{ds}$ < P+. Our device is also electrically bipolar: in fact, during a voltage polarity sweep restricted to positive values +$V_{ds}$ (or negative, -$V_{ds}$), the step of the accumulated charge Δq or $ΔI_H$ appears only for the first sweep unless an opposite voltage -$V_{ds}$ (+$V_{ds}$) is applied to the device (see Figure S1c).
In each case, the first cycle shows a current transition from state 0 to 1 at the inflexion point but for the second cycle and above, it is necessary to sweep to the opposite erase area (of opposite sign in $V_{ds}$) in order to recover the hysteresis effect during a second cycle. If this is not the case, there is no transition in the device and it remains in the 0 state for the rest of the cycles, as shown in the Figure S2b.
Note that the hysteresis position and amplitude remain quite similar when the bias sweep rate is tuned from 20mV/s to 2V/s.

Within our bipolar memristor device, it is possible to block the current in the 0 state or the 1 state by sweeping $V_{ds}$ carefully between the different regions

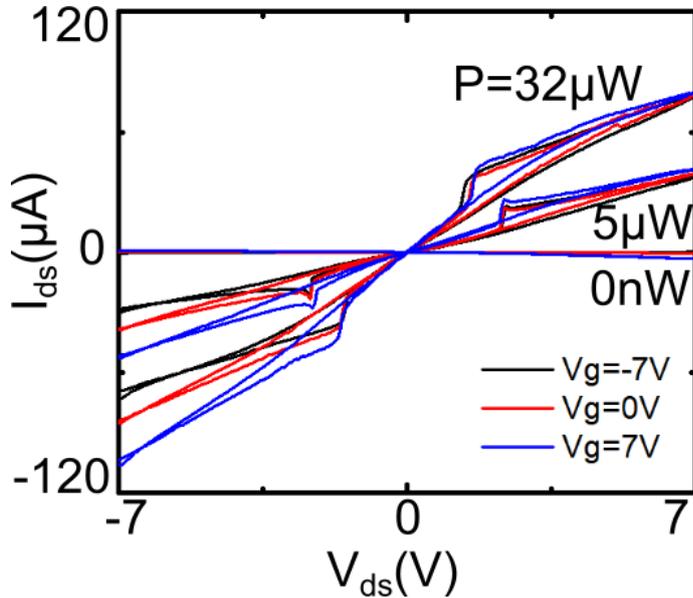

**Figure S1a Mesure of $I_{ds}(V_{ds})$ for different $V_g$ and laser power.** The gate voltage does not induce any notable modification of the hysteresis position or amplitude even if a decrease of the resistance value is seen with higher gate voltage in the current-voltage dependence

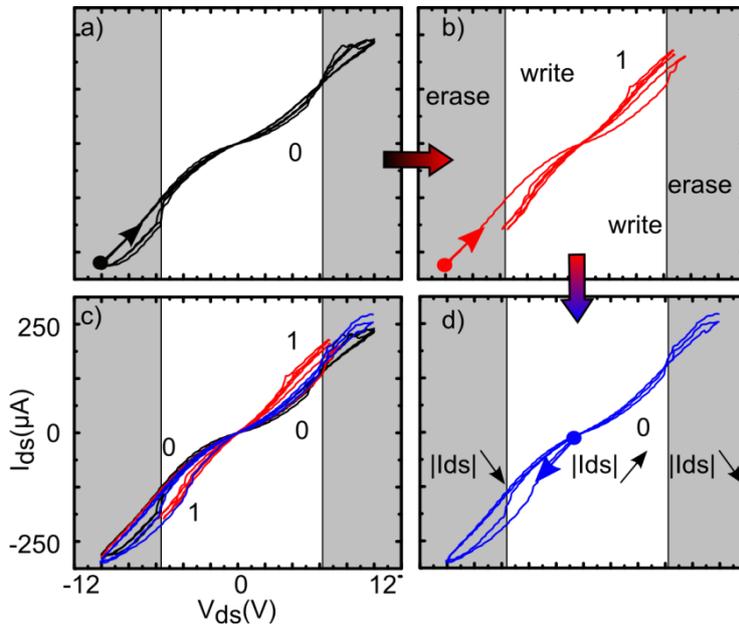

**Figure S1b Memory effect in the MoS$_2$ device with state 0 and 1.** a) $I_{ds}(V_{ds})$ curve with a starting point at $V_{ds}$=-10V and sweep from -10 to 10V. b) $I_{ds}(V_{ds})$ curve with a starting point at $V_{ds}$=-10V and sweep from -5.8 to 5.8V. d) $I_{ds}(V_{ds})$ curve with a starting point at $V_{ds}$=0V and sweep from -10 to 10V. d) The three measurements are taken in the order: black then red then blue.

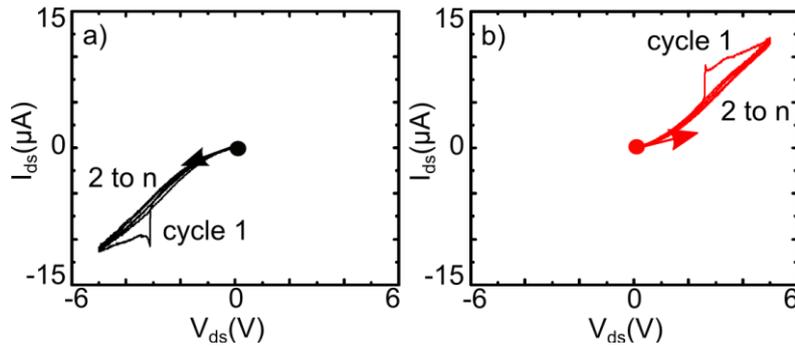

**Figure S1c Memory effect for only one cycle when V$_{ds}$ sign does not change.** a) I$_{ds}$(V$_{ds}$) curve with a starting point at V$_{ds}$=0V and few back and forth sweeps from 0 to -10V. b) I$_{ds}$(V$_{ds}$) curve with a starting point at V$_{ds}$=0V and few back and forth sweeps from 0 to 10V.

## S2: Others nanomechanical memrisitive device

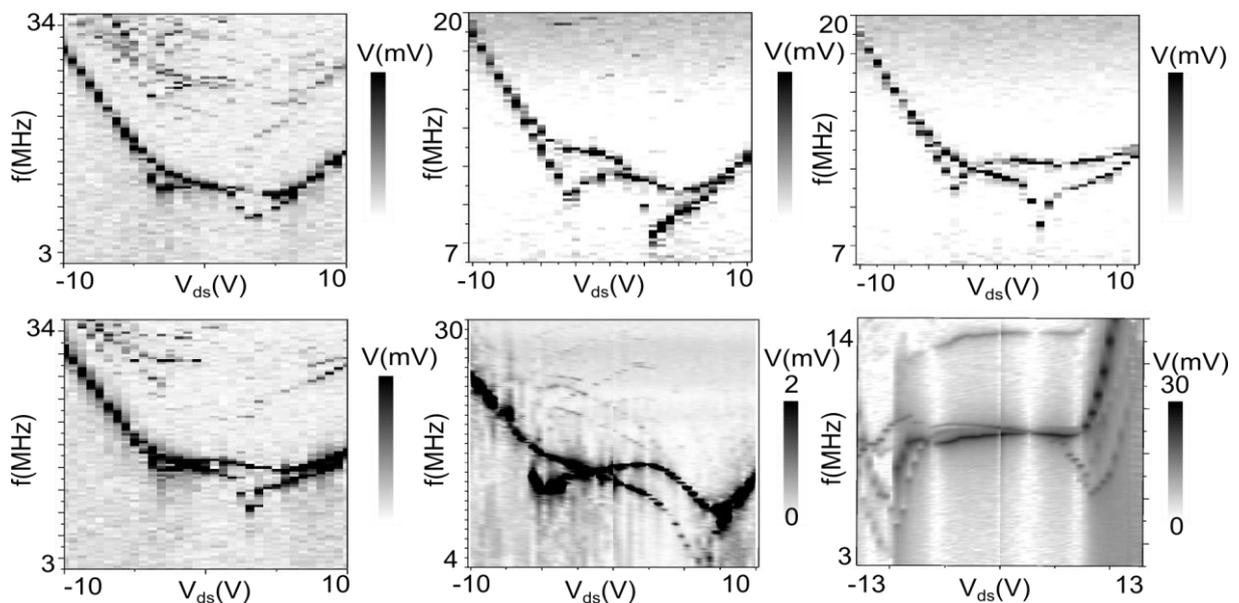

**Figure S2a Memory effect for only one cycle when V$_{ds}$ sign does not change.** a) I$_{ds}$(V$_{ds}$) curve with a starting

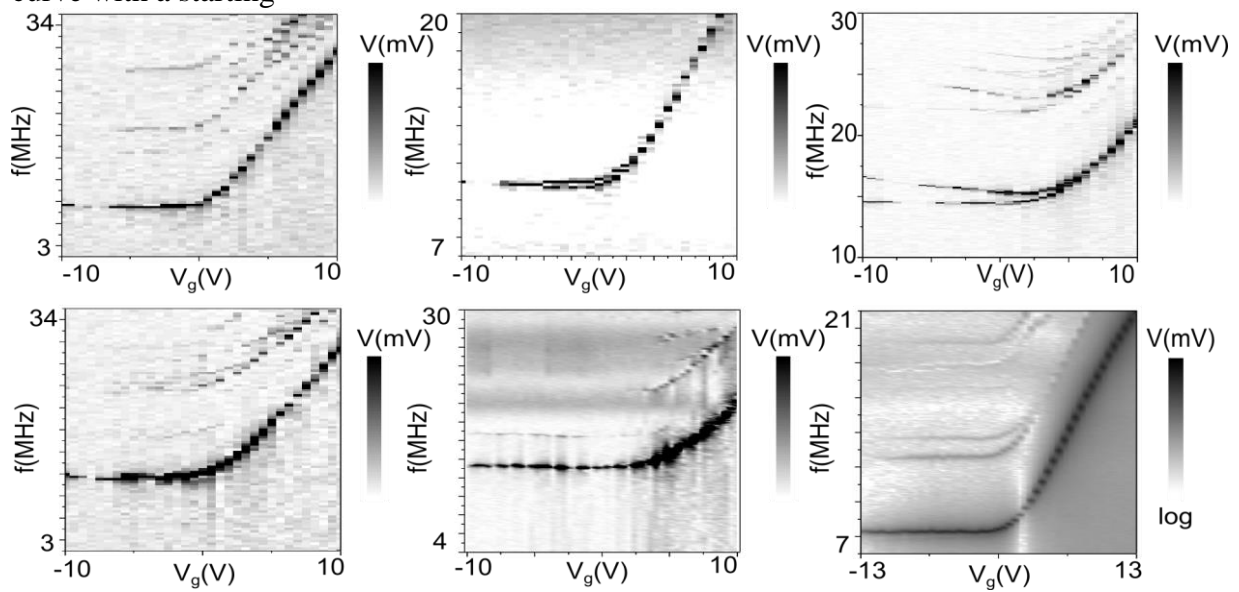

**Figure S2b Memory effect for only one cycle when $V_{ds}$ sign does not change.** a) $I_{ds}(V_{ds})$ curve with a starting

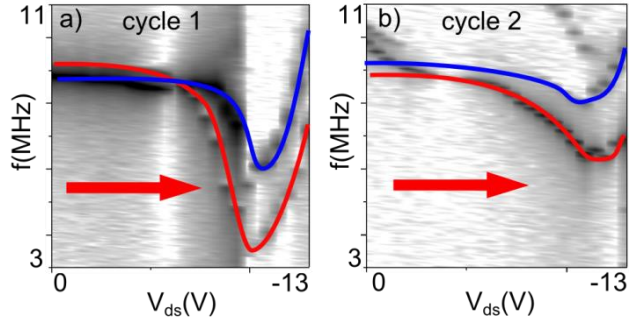

**Figure S2c Memory effect for only one cycle when $V_{ds}$ sign does not change.** a) $I_{ds}(V_{ds})$ curve with a starting

## S3: Fully clamped drum resonator model
[5–10]

### Static deflection of a circular membrane

Considering a very thin plate, fully clamped, we work in the membrane limit and neglect the flexural rigidity of our 2D resonator. In circular coordinate, the static shape w(r) of a doubly clamped drum resonator is assumed to be a parabolic shape;

$$w(r) = z\left(1 - \frac{r^2}{R^2}\right) \quad (2)$$

With z the deflection at r=0 and R the radius of the membrane
The elastic energy $U_{el}$ is the sum of the bending energy $U_b$ and of the restoring potential energy $U_T$ which comes from its tension. In our case, we have already demonstrated our membrane to be in the tension limit. We only considered the strain energy per unit volume, defined as [8,11,12];

$$dU_{el} \approx dU_T = \sum_{i=r,\theta}\sum_{j=r,\theta} \sigma_{ij}\varepsilon_{ij}dV \quad (3)$$

The approximation with a drum has a central symmetry. We consider w the lateral deflection, u the radial displacement and $\varepsilon_{r,\theta}$ the radial and angular strain and we can write

$$\varepsilon_{rr}(r)dr = \varepsilon_r(r)dr = \frac{\partial u}{\partial r}.dr + \frac{1}{2}\left(\frac{\partial w}{\partial r}\right)^2.dr \text{ and } \varepsilon_{\theta\theta}(r) = \varepsilon_\theta$$
$$= \frac{u}{r} \text{ and } \varepsilon_{r\theta} = 0 \quad (4)$$

Here, $\partial u/\partial r$ is the radial elongation. In our case we replace this term by the initial radial strain-stress duo $\varepsilon_0$-$\sigma_0$ which is consider to be uniform along the sample. $2\pi R.\varepsilon_0 = \iint \varepsilon_0 r dr d\theta$. From

Hooke's law, if the tangential strain is assumed to be 0, then we have $\sigma_r - \sigma_0 = \frac{Et}{(1-v^2)}\varepsilon_r$ with v the Poisson's ratio and E the Young Modulus. Finally, it gives;

$$U_{el} = \frac{1}{2}\frac{Et}{(1-v^2)}\int_0^{2\pi}\int_0^R \left(\varepsilon_0 + \frac{1}{2}\left(\frac{\partial w}{\partial r}\right)^2\right)^2 r\,dr\,d\theta \quad (5)$$

In other terms;

$$U_{el} = \frac{2\pi Etz^4}{3(1-v^2)R^2} + \frac{\pi E\varepsilon_0 tz^2}{(1-v^2)} + \frac{\pi Et\varepsilon_0^2 R^2}{2(1-v^2)} \quad (6)$$

$$U_{el} = \frac{\pi Et}{(1-v^2)} \cdot \left(\frac{2z^4}{3.R^2} + \varepsilon_0 z^2 + \frac{\varepsilon_0^2 R^2}{2}\right) \quad (7)$$

$$U_{el} = \frac{\pi Et}{(1-v^2)} \cdot z^2 \left(\frac{2z^2}{3.R^2} + \varepsilon_0\right) + \frac{\pi Et}{(1-v^2)} \cdot \frac{\varepsilon_0^2 R^2}{2} \quad (8)$$

The second term is the initial energy $U_{init}$ resulting from the build-in stress; $U_{init} = 1/2 \int \sigma_0 \varepsilon_0 r\,dr\,d\theta$.

If we define the total stress T for a small angle dθ applied to the membrane as the sum of building stress and the stress due to the deflection of the membrane for the same angle, we have [8];

$$T(z) = T_0 + T_D(z) = T_0 + \varepsilon_D(z)\frac{Et}{(1-v^2)} \quad (9)$$

$$T(z) = T_0 + \frac{Et}{(1-v^2)}\frac{1}{R}\int_0^R \frac{1}{2}\cdot\left(\frac{\partial w}{\partial r}\right)^2 dr = T_0 + \frac{2}{3}\frac{Et}{(1-v^2)}\frac{z^2}{R^2} \quad (10)$$

If we integrate T over all angle $\int_0^{2\pi} d\theta$, an multiply by $z^2.t/2$, we retrieve the first term in $U_{el}$. In other term;

$$U_{el} = \frac{1}{2}.z^2.2\pi.T(z).+U_{init} \quad (11)$$

This notation will be important for the dynamical description of our membrane.

We want also to noticed that, with similar notation and with a central symmetry, the Bending energy is defined as [9];

$$U_B = \frac{1}{24}\frac{Et^3}{(1-v^2)}\int_0^{2\pi}\int_0^R \left(\frac{\partial^2 w}{\partial r^2} + \frac{1}{r}\frac{\partial w}{\partial r} + \frac{1}{r^2}\frac{\partial^2 w}{\partial \theta^2}\right)^2 rdrd\theta \quad (12)$$

It is also interesting to describe the electrostatic energy $U_C = 1/2 C_g V_g^2$ and develop the capacitance term for small deflection $C_g = C_0 + C'z + C''z^2/2$ where $C' = dC_g/dz$ et $C'' = d^2C_g/dz^2$ Then the electrostatic force is

$$U_C = \frac{1}{2}C_0 V_g^2 + \frac{1}{2}C'zV_g^2 + \frac{1}{4}C''z^2V_g^2 = \frac{1}{2}V_g^2 \cdot \epsilon_0 \cdot \iint \left(\frac{1}{d^1} + \frac{w}{d^2} + \frac{w^2}{d^3}\right) rdrd\theta$$

$$= \frac{1}{2}\left(\frac{\epsilon_0 \pi R^2}{d^1}\right)V_g^2 + \frac{1}{2}\left(\frac{\epsilon_0 \pi R^2}{2d^2}\right)zV_g^2 + \frac{1}{4}\left(\frac{\epsilon_0 2\pi R^2}{3d^3}\right)z^2 V_g^2 \quad (13)$$

We can notice the minus sign appearing in the resulting Force $F_C = dU_C/dz = -1/2\, dC_g/dz\, V_g^2$ because we have to apply a fix voltage $V_g$ with an external battery [13,14] and we must consider the virtual work done on the charging source $U_b = -C_g V_g^2$ to the energy stored in the capacitor. It can also be deduced by derivate the energy, working at constant charge, $U_C = 1/2\, q^2/C_g$.

The equilibrium position z is defined by the minimum of energy or $\partial(U_{el} + U_C)/\partial z = 0$

$$\frac{8\pi E t}{3(1-v^2)R^2}z^3 + \left(\frac{2\pi E \varepsilon_0 t}{(1-v^2)} - \frac{1}{2}C''V_g^2\right)z - \frac{1}{2}C'V_g^2 = 0 \quad (14)$$

It is a polynomial equation of the form $z^3 + az + b = 0$ with solutions defined by the Cardan method;

$$z = \left((-b-\sqrt{\Delta})/2\right)^{1/3} + \left((-b+\sqrt{\Delta})/2\right)^{1/3} \text{ and with } \Delta = b^2 + 4a^3/27 \quad (15)$$

It defined a relation between z and $V_g$. This equation describes our system only when a deflection z (and w) are no null. It is valid in a regime where the capacitive force $V_g$ is not negligible (high $V_g$) or when this initial strain itself induced some deflection. At low $V_g$ and in a built-in strain in the plane $\varepsilon_r = \partial u/\partial r$, we are in a regime with no deflection (w=z=0) and $U_{el} = \pi E t \varepsilon_0 R^2/2(1-v^2)$. A modification of the built-in strain will not modify the equilibrium position and will not create a deflection. This has to be opposite to a vibration which is an oscillating deflection (w≠0) and the build-in strain can induce a vibration frequency modification. In other words, a planar modification of the build-in strain can explains our measurements where we obtained a hysterical softening of the vibration and no deflection of the membrane at the same time.

**Dynamical deflection of a circular membrane**

In this section we will transform w→w+$w_{AC}$ and $F_C$→$F_C$+$F_{AC}$. However, we will also assume $T_M(z)$ and $\sigma_R$ to be constant because the contribution of a small oscillation around the equilibrium position to the stress will converge to 0 in a symmetric harmonic potential. It is unambiguous from equation 10 that our system is equivalent to a spring which can vibrate. The main difference consists in the spring constant itself which depend on the deflection amplitude.

The motion equation in this case, is [10];

$$\rho \frac{\partial^2 w}{\partial t^2} = -\kappa \nabla^4 w + T\nabla^2 w + \frac{F_C}{\pi R^2} \tag{16}$$

In polar coordinates, $\nabla^2$ is $\frac{1}{r}\frac{\partial}{\partial r} + \frac{\partial^2}{\partial r^2}$, $\rho$ is the 2D mass density and $\kappa$ is the bending stiffness. We know the bending term to be negligible in our case[1]. As expected, this equation gives the same result for z, $T_M$ and $F_C$ and confirm our description of the harmonic oscillator submitted to a force $F_C$. Concerning the dynamic counterpart, the mode shape of a circular membrane can be expressed using $J_0$, the 0th-order Bessel function of the first kind [15]. Considering the circumference-clamped boundary conditions, the mode shape of the resonance is:

$$w_{AC}(r) = \sum_{n=0}^{\infty} \delta z_n \cdot J_0\left(\lambda_n \frac{r}{R}\right) e^{-i\omega_n t} \tag{17}$$

Where $\delta z_n$ is the complex amplitude dynamical elongation at the center. For the fundamental mode $\lambda_0 = 2.405$.

During the vibration of this membrane, the frequency of oscillation $f_0$ can be described by a harmonic oscillator with a spring constant k and an effective mass m;

$$f_0 = \frac{1}{2\pi}\sqrt{\frac{k}{m_{eff}}} \tag{18}$$

The spring constant can be obtained by second order differentiation of the total potential energy, with the elastic $U_{el}$ and electrostatic contribution $U_c$ and the effective mass can be obtained from the Kinetic energy $U_{cin}$ expression

$$U_{cin} = \frac{1}{2}m_{eff}\delta z^2 \tag{19}$$

And also

$$U_{cin} = \frac{\rho}{2}\int_0^{2\pi}\int_0^R w(r)^2 r dr d\theta = 2\pi\frac{\rho}{2}\int_0^R \left(\delta z \cdot J_0\left(2.405\frac{r}{R}\right)\right)^2 r dr \tag{20}$$

$$U_{cin} = 0.136\pi\rho R^2 \delta z^2$$

It gives:

$$m_{eff} = 0.271\rho\pi R^2 \tag{21}$$

The spring constant is defined as $k = \partial^2(U_{el} + U_C)/\partial z^2$, if we take the equation 10 and 13;

$$U_{el} = \frac{1}{2}.z^2.2\pi.T(z) + U_{init} \tag{22}$$

From this equation and equation 13, we will derive the spring constant as
$k = T_0 + \frac{12}{3}\frac{Et}{(1-\nu^2)}\frac{z^2}{R^2} - \frac{1}{2}C''V_g^2 = T_2$ with a renormalized tension $T_2$, compare to T, which take into account the z dependence of T and the capacitive softening. We can write the resonant frequency of the fundamental mode as

$$f_0 = \frac{2.405}{2\pi R}\sqrt{\frac{T_2}{\rho}} \tag{23}$$

$$f_0 = \frac{1}{2\pi}\sqrt{\frac{2.405^2\left(T_0 + \frac{12}{3}\frac{E}{(1-\nu^2)}\frac{z^2}{R^2}\right)}{R^2\rho} - \frac{\epsilon_0 V_g^2}{0{,}271 x 3. d^3. \rho}} \tag{24}$$

If we integrate in this equation, the bending of the membrane, define in the equation 11, we obtain an expression as;

$$f_0 = \frac{1}{2\pi}\sqrt{2.405^4\frac{Et^3}{12(1-\nu^2)}\frac{1}{R^4\rho} + 2.405^2\frac{T_0}{R^2\rho} + 2.405^2\frac{12}{3}\frac{Et}{(1-\nu^2)}\frac{z^2}{R^4\rho} - 1.23\frac{\epsilon_0 V_g^2}{d^3\rho}} \tag{25}$$

And z is defined by the equation 14.

### S4: Mechanical hardening and softening

The list of potential candidates related to a shift of a nanoresonator frequency are quite long; as a dopant absorption, an electrostatic force, a chemical potential variation, a chemical reaction, a photothermal effect, a joule heating, a slip of the $MoS_2$ at the clamped side, a defect diffusion, all nonlinear mechanical behaviors, an internal strain variation due to a crystal phase transition, a polarized piezoelectric effect. We investigate and explain our unique behaviors as the strong mode softening under the light of these different scenarios.

For our knowledge, in this list, only a few scenarios explain an in-plane strain shift; a phase transition, the piezoelectric effect and a dielectric polarization force induced by an in plane ununiformed polarization P or electric field [16]. But only the phase transition can be applied to our case and is strong enough to explain the 2.7 reduction in mechanical frequency.

### Mass deposition;

Atomically thin mechanical resonators are well known to be mass sensitive but the $MoS_2$ is relatively heavy (160 Daltons per atomic mesh). The molecular absorption hardly explains our frequency reduction. It corresponds to the absorption and desorption of more or less 30-56 layers of water molecules, for example, on top of the $MoS_2$ [17]. Consequently, the hysteretic softening is related to the strain itself. Absorption-desorption is not excluded by a strong decrease of strain but considering that ions incorporation is usually hardening a material and is not reversible in a vacuum environment, it contradicts our observation.

In detail; We can consider the case of additional dopant absorption in or on the semiconductor. The sample is in high vacuum (P~$1.10^{-5}$mBar) but molecules like water or oxygen are still present in the cryostat and can affect the doping. We can put this scenario on one side if we consider the mechanical motion and reflection measurements. We consider a resonator of frequency $f=1/2\pi.(k/m)^{0.5}$, with m the effective mass of the resonator and k the spring constant. With mechanical motion detection, we are sensible to additional mass like within a chemical reaction or a layer deposition of absorbent. The dopant absorption, diffusion or desorption along the sample will modify f by $\delta f$ and the additional mass will be $\delta m=2m.\delta f/f$ [18]. If the frequency is reduced by 2 or 2.7 like in our mechanical memory measurements, it corresponds to an additional mass equivalent to 4-7 times the $MoS_2$ mass. It has been shown previously the possibility to fully recover a nanoresonators with layers of atoms[17] but let consider this case. The mass of a mesh of $MoS_2$ $M_{MoS2}=M_{Mo}+ 2M_S=160$ Daltons with $M_{Mo}=96$ Daltons and $M_S=32$ Daltons. The mass of a $H_2O$ or $0_2$ molecules are 18 Daltons and 16 Daltons respectively. It means our frequency reduction corresponds to the absorption and desorption of more or less 30-56 molecules per mesh of $MoS_2$ and more than 30-56 layers of molecules on top of the $MoS_2$ [17]. It is unlikely that this phenomenon is at the origin of our behaviors.

**Capacitive softening**

Mechanical hardening or softening are often observed in similar 2D devices or nanotube [19,20] due to the sensitivity of these materials to external inputs. Indeed softening is related to the geometrical ratio between the membrane thickness t and the distance d to the gate electrode [21], and is improve together with the electrostatic coupling, but a comparable strong softening was only observed in graphene resonators[5] and attributed to a strong and attractive electrostatic force with a very close gate (<150nm) [21]. Such a large reduction is expected when the capacitive force becomes comparable to the restoring force of the resonator. When the two forces are equal, the frequency drops to zero and the resonator collapses against the counter electrode. In our sample with large d, around 660nm, this capacitive softening is much less efficient. In fact, this mechanical hysteretic softening does not depend on the gate voltage, as shown in Figure 2c and at first glance is not related to capacitive softening.

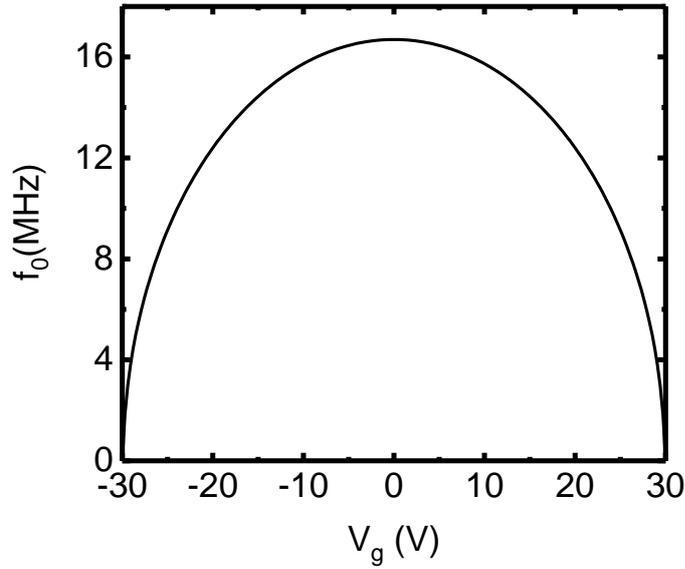

**Figure S4a Capacitive softening simulation of the suspended MoS$_2$.** We plot the capacitive softening term for the sample of Figure 2 in function of $V_G$ (neglecting all others terms except the internal tension in equation 24). The memristive effect decrease the frequency at $V_H$~2.5V which is an order of magnitude more efficient than the capacitive term. Also the mechanical softening effect does not show any dependence in $V_g$ but only in $V_{ds}$ which is in contradiction with the capacitive softening term.

### **Heating;**
Also, despite the threshold of the memristive effect appearing at the same power $P_H=I_H.V_H$ independently from the photoconductivity (in Figure 2a and in SI), all the sample heating hypothesis can be easily ruled out. The photothermal effect induce by a laser of 10μW is negligible on our sample, as we have measure previously with Raman spectroscopy [1], as well as the thermal dilatation. The Joule heating very small and heat the sample by few tens of degrees for an applied $V_H$ and $I_H$, as shown in Figure 2b, and based on our COMSOL simulation (see SI). The temperature does not increase consequently, more than 177 K, if we remove the thermal coupling to SiO$_2$ pillars. However, since the temperature improve and higher $V_{ds}$ polarization bring an out-of-equilibrium situation with hot electrons in the system, it certainly promotes the $S_V$ diffusion or $S_V$ clustering or a phase transition but the photocurrent and the mode softening are uncorrelated to any heating effect

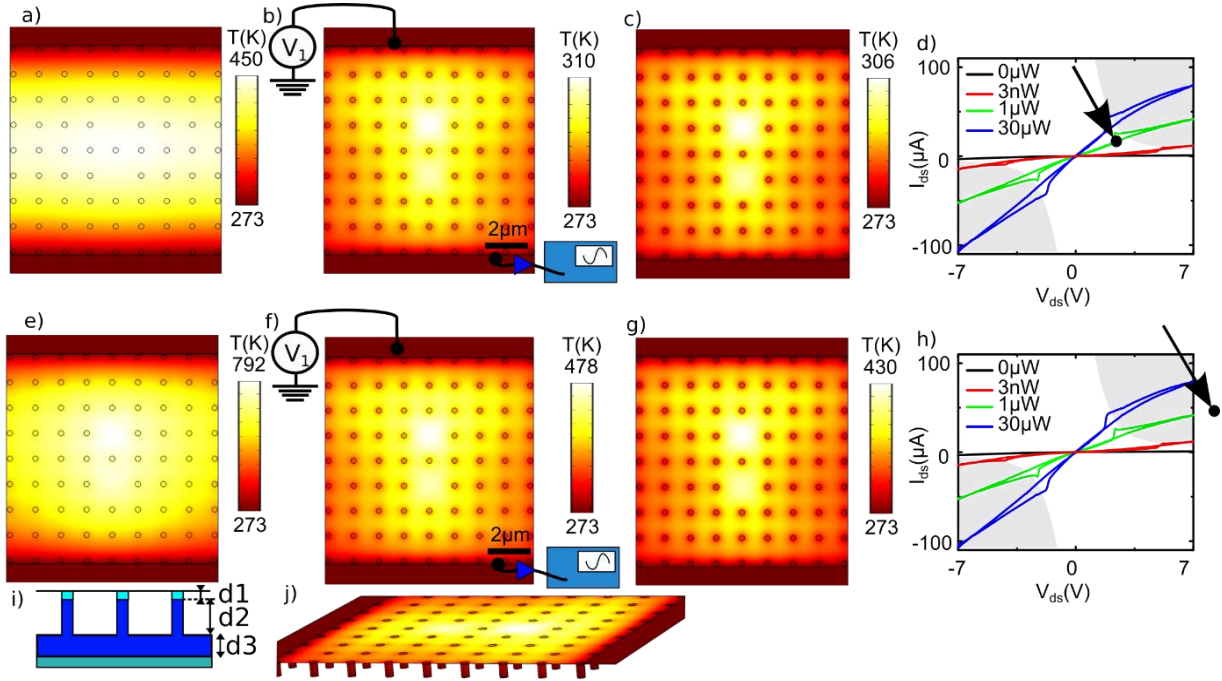

**Figure S4b Joule heating in the suspended MoS$_2$.** It is possible to heat the MoS$_2$ by Joule heating when a voltage is applied. a) b) and c) Joule heating, obtained with COMSOL simulation at $V_{ds}$=2.4V and $I_{ds}$=16µA as shown by the point in d) It corresponds to the jump in current in the memristive device. e) , f) and g) the Joule heating simulation at $V_{ds}$=10V and $I_{ds}$=42µA. i) a schematic which shows the MoS$_2$ and the three different parts of SiO$_2$ with thickness $d_1$=50nm, $d_2$=520, $d_3$=150nm and the Si substrate. j) the system under test with another point of view in order to see the pillars. We have changed the thermal coupling with the SiO$_2$ pillars by inserting a part d1 in the pillars structure with a reduced thermal conductivity $k_{d1}$. In a) it is 0W/m/K, in b) and f) it is $k_{d1}$=$k_{SiO2}$/10, in e) it is $k_{d1}$=$k_{SiO2}$/100 and in c) and g) $k_{d1}$=$k_{SiO2}$ with a perfect thermal coupling.

In the simulation, the MoS$_2$ is suspended with a thermal conductivity of $k_{MoS2}$=50W/m/K, the SiO$_2$ pillars has a thermal conductivity of $k_{SiO2}$=1.4W/m/K. It is possible to observe at the charge accumulation point that the power P=$I_{ds}$.$V_{ds}$ only improve the temperature by few tens of degree. Even if the MoS$_2$ is completely decoupled from the pillars, the temperature only increase by 177 K which is not enough to explain a phase transition by Joule heating. We have also conducted some simulation with higher current and voltage and the temperature begin to rise if the decoupling to the SiO$_2$ pillars is quasi total. During the simulation the doping of the MoS$_2$ what convoluted with a Gaussian of 6µm of diameter in order to simulate the laser doping and placed at x=0µm and y =-1.26µm, in the center of the larger membrane where the Joule effect is the strongest. The result is not strongly different than the case with a uniform doping. For the MoS2 conductivity, only the voltage and the resulting current where fixed.

### S$_V$ diffusion along the membrane
In contrast, S$_V$ centers were assumed to stress the material locally[22], which is in contradiction with our observations. Additionally, our monocrystalline device exhibits this softening at the location of the laser excitation—far from the edges. If electromigration and Schottky barriers had a role to play in our samples, it would be conceptually different from ref[23], where ΔI$_H$ decreased and the material degraded at the edges.

### Membrane slipping
If we consider a membrane slipping on the anchoring, it can produce a mechanical relaxation of the build in strain but not a huge bipolar reversible effect. Our mechanical memristive behavior is not due to a slipping of the membrane

### Piezoelectricity
Most of our device present some increase of frequency for high $|V_{ds}|$ and we assume it is a piezoelectric hardening of our suspended device. The in plane piezoelectric coefficient is $d_{11}$=2.5-4 pm/V and the out-of-plane ferroelectricity coefficient was recently measured to be $d_{33}$=1.3 pm/V (Two-dimensional materials with piezoelectric and ferroelectric functionalities) and at $V_{ds}$=10V the additional strain is . If we consider this strain to be effective along the whole suspended $MoS_2$ and not only in our membrane under measurement, we will have to consider the stress applied on the membrane to be ?? which corresponds quite well to our data. Also, piezoelectricity presents a hysteresis in a strongly polarized material as a ferroelectric. 2H-$MoS_2$ is centrosymmetric and unfavorable to ferroelectricity and any intrinsic polarization. Even the 1T' phase, which presents a small out of plane polarization of $0.18\mu C/cm^2$ and a $T_C$ at (ref Emergence of Ferroelectricity at a Metal-Semiconductor Transition in a 1T Monolayer of MoS2 and http://link.aps.org/supplemental/10.1103/PhysRevLett.112.157601) does not explain this polarization. In a non-uniform distribution of charge as in a polarized dielectric material non uniformly doped by a laser, the associated current density corresponds to the movement of dipole moments density P as $J = \partial P/\partial t$. Since the dopants diffusion and current contribution is mostly associated to $S_V$ vacancies and very long time scale in our sample, it is difficult to link any polarization to the relatively fast memristive behavior.

### Dielectric force
Until now we have only consider electrostatic force to be capacitive coupling between two parallel plates with a uniform charge density. In reality, the semiconducting 2D material shows a high doping signature induced by the laser illumination and the charge density on the $MoS_2$ cannot be uniform anymore. It seems reasonable to approximate this density with a 2D Gaussian distribution. We consider a dielectric in a non-uniform electric field. The energy of an electrostatic system over a volume V is [24]

$$U_D = \frac{1}{2}\int_V^0 \boldsymbol{D}.\boldsymbol{E}dv = \frac{1}{2}\int_V^0 \epsilon_0\epsilon_R E^2 dv \qquad (26)$$

It is possible to derive a Force $F_D$ base on a dipole approximation. The force applied on a unique dipole with a small size compare to the rest of the system is;
$$F_{D1} = \boldsymbol{p}.\boldsymbol{\nabla E} \qquad (27)$$

A dielectric material can be approximate by multiple non interacting dipoles with a certain density and a corresponding macroscopic polarization density P. Then, if we consider a static electric field ($\boldsymbol{\nabla} \times \boldsymbol{E} = 0$). the equation for a dielectric become
$$F_D = \boldsymbol{P}.\boldsymbol{\nabla E} = \epsilon_0\chi\boldsymbol{E}.\boldsymbol{\nabla E} = \frac{1}{2}\epsilon_0[(\epsilon_R - 1)\nabla EE] \qquad (28)$$

$$F_D = \boldsymbol{P}.\boldsymbol{\nabla E} = \epsilon_0\chi\boldsymbol{E}.\boldsymbol{\nabla E} = \frac{1}{2}\epsilon_0[\nabla((\epsilon_R - 1)EE) - \boldsymbol{E}.\boldsymbol{E}\nabla(\epsilon_R - 1)] \qquad (29)$$

**E.E** is a tensor defined as

$$\mathbf{EE} = \begin{pmatrix} E_x^2 & E_xE_y & E_xE_z \\ E_xE_y & E_y^2 & E_yE_z \\ E_xE_z & E_yE_z & E_z^2 \end{pmatrix} \quad (30)$$

$$\nabla = \begin{pmatrix} \partial/\partial x \\ \partial/\partial y \\ \partial/\partial z \end{pmatrix} \quad (31)$$

The term which is in plane can be consider as a pressure applied on a dielectric slab insert in the capacitor, of polarization density P, thickness t and width W. It represents for example some moving dopants along the MoS$_2$. The Force on the polarized plate, along x, would be

$$F \sim \frac{1}{2}\epsilon_0(\epsilon_R - 1)\left(\frac{V}{d}\right)^2 Wt \quad (32)$$

At the end, we have a strong analogy of this force with the capacitive force describe earlier. And theappearance of this force in the

## S5: Relation between the reflectance and the cavity length.

In order to optimize the optical cavity length in between the MoS$_2$ and the substrate, we have defined the reflectance R on a laser at 633nm on the sample for both samples geometries. Sample A with d=520nm and d$_{SiO2}$=150nm and sample B with d=440nm and d$_{SiO2}$=220nm

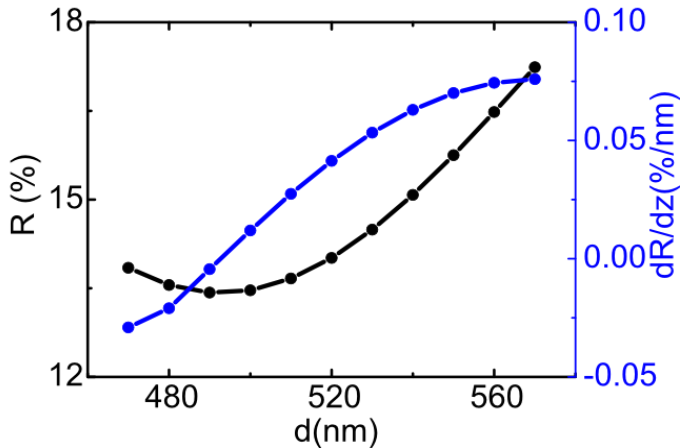

**Figure S5 Relation between the reflectivity R and the cavity length d.** We have plot the reflectance of the laser in function of the distance d between the MoS$_2$ and the SiO$_2$. In blue we have also plot the derivative dR/dz which is directly proportional to our mechanical signal amplitude. Here the SiO$_2$ thickness is 150nm and the MoS$_2$ is a monolayer flake. In this device, the SiO$_2$ thickness is 520nm with dR/dz at ~0.04%/nm.

## S6: Memristor energy

The energy of a memristor during a step of current I$_{ds}$ in the I$_{ds}$-V$_{ds}$ curve is equivalent to a resistance energy with a times dependence of M(q(t)):

$$U_M = \frac{V_{ds}^2}{2}\int_{\Delta t}^{0}\frac{1}{M(q(t))}dt = \frac{V_{ds}^2}{2}\int_{\Delta q}^{0}\frac{1}{I_{ds}.M(q)}dq = \frac{V_{ds}^2}{2}\frac{\Delta q}{V_{ds}} = \frac{1}{2}V_{ds}\Delta q \quad (33)$$

Since the energy does not depend on the deflection z and does not introduce a planar strain in the membrane, it will not directly intervene in the motion equation. But, for a fix $V_g$, and the gate charge equilibrium related to an external battery the capacitance energy will change by
$U_C = 1/2\,(q + \Delta q)^2/C_g = 1/2\,q^2/C_g + q\Delta q/C_g$

In ref [25], they consider the magnetic flux create by the moving dopant in the first memristor device [3]. This magnetic flux is no null but $10^5$ times lower than the magnetic flux created by the current charge. We will do the same in our system and neglect all magnetic force in our system, especially since the integration over a 2D space will gives lower values of the magnetic field.

## S7: Persistent current measurements

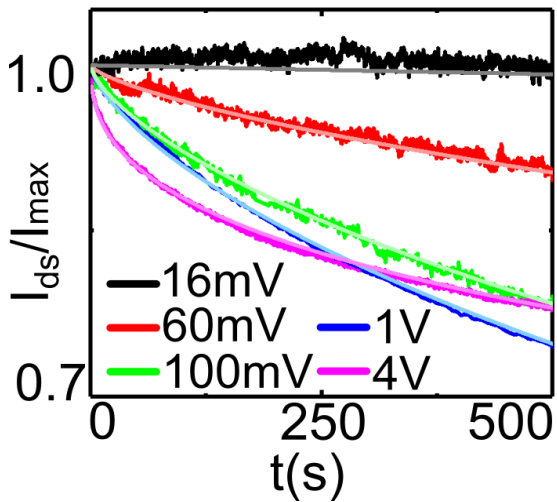

**Figure S7** Time relaxation time of photocurrent at different $V_{ds}$

One can clearly notice that the relaxation time of the system strongly depends on the applied voltage. Thus, several regimes can be extracted: (i) at $V_{ds} = 16$ mV, the current was almost fixed at the same value when the excitation was OFF; (ii) around 60 mV, the long lasting relaxation time can be fitted by almost a single exponential function; (iii) between 100 mV and 1 V, the decay of the current can be fitted with good accuracy by means of a double exponential function, corresponding to a system with two types of traps or dynamics ($I(t) = I_1 e^{-t/\tau_1} + I_2 e^{-t/\tau_2}$)[26,27]; (iv) beyond $V_H$ (at $V_{ds} = 4$V), we detect an additional contribution with a faster decay time $I(t) = I_1 e^{-t/\tau_1} + I_2 e^{-t/\tau_2} + \Delta I e^{-t/\tau_3}$, where $\tau_1$ and $\tau_2$ correspond to the long decay times in the order of 269 ± 48 s and 8815 ± 690 s, respectively, while $\tau_3$ is relative to the fast decay time with a value around 20 s. Note that these PPC timescales were in accordance with previous studies[27,28]. Furthermore, the contributions of $\Delta I$, which is related to the memristive effect remained very low (4% of the total current) and difficult to analyze. To overcome this issue, we measured the I-V curve when the excitation was ON: we followed the dynamics of $I_{ds}$ when the voltage was quickly swept from 0 V to a certain $V_{ds}$ and then fixed at this value. After almost twelve seconds, the photocurrent generation $I_{photo}$ dominate the measurements and is always positive for an extremely long time, independently from $V_H$. However, during the first ten seconds of the measurements, in Figure 4i, we observed again the appearance of a current assimilated to $\Delta I$ with the same dynamic and $V_H$ dependence.

We have fit with exponential decay function with multiple decay time of the type $I(t) = I_1 e^{-t/\tau_1} + I_2 e^{-t/\tau_2} + I_3 e^{-t/\tau_3}$. Except the measurement at 16mV, most of the measurements where much longer than 500s and the fit for the long time scale (~8000s) is quite accurate.

$\tau_1, \tau_2, \tau_3$ are in seconds and the current is normalized at 1 for t=0s
At $V_{ds}$=16mV , $I(t) = 1e^{-t/60000}$.
At $V_{ds}$=60mV , $I(t)=0.04 e^{-t/287}+0.95 e^{-t/8300}$.
At $V_{ds}$=100mV, $I(t)=0.184e^{-t/272}+0.802e^{-t/8921}$.
At $V_{ds}$=1V      , $I(t)=0.248e^{-t/328}+0.737e^{-t/8180}$.
At $V_{ds}$=4V      , $I(t)=0.125e^{-t/191}+0.81e^{-t/9922}+0.04e^{-t/19}$.

At $V_{ds}$=60mV, the contribution of $I_1$ is only 4% of the total current and represents 18.6 ± 5.3% for higher $V_{ds}$. The contribution to the short time $I_3$, at $V_{ds}$ = 4V, is also very small at 4%.

**S8: 1T' phase signature in our MoS$_2$ with Raman and photoemission spectroscopy**

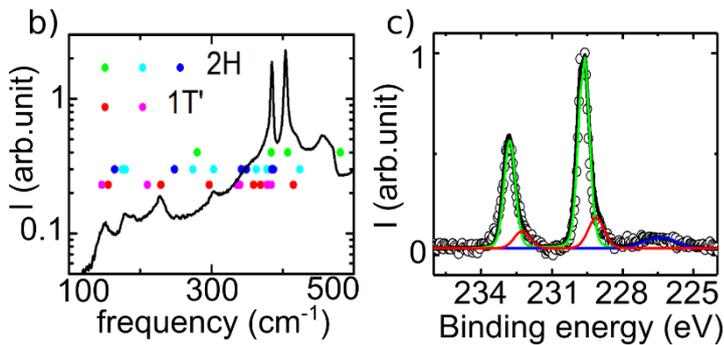

**Figure S8** b) Raman spectroscopy of our MoS$_2$ at ambient conditions with the two main peaks of 2H-MoS$_2$ and small peak contributions from the 1T' phase. c) Photoemission signal of the MoS$_2$ with a main green peak contribution for the 2H phase and an additional red peak attributed to the 1T' phase. Both measurements indicated the presence of a few % of this phase in the material.

**S9: Writing/ erasing processes and 1T' phase evidence in our MoS$_2$**

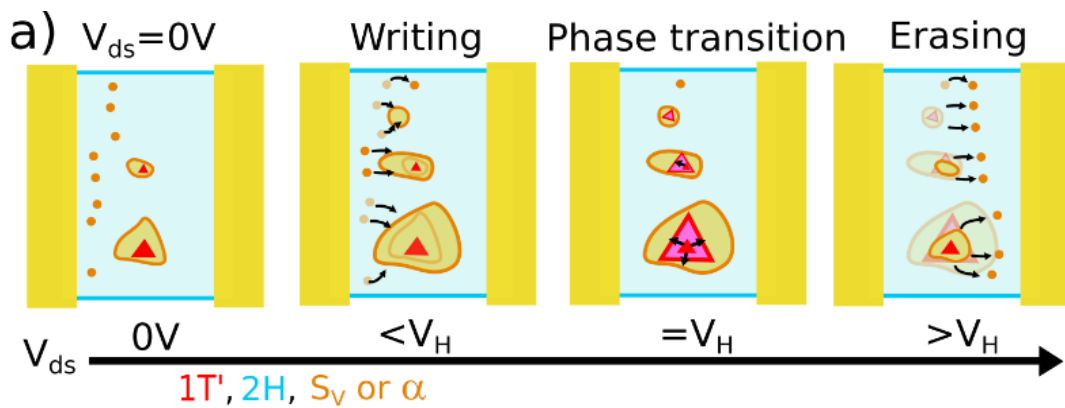

**Figure S9** a) Schematic of the writing/erasing process for positive $V_{ds}$ values. Below $V_H$, $S_V$ centers migrated to the right and accumulated in or created α clusters. At $V_H$, the α phase promoted the nucleation of the phase 1T in the 2H phase. Beyond $V_H$, the field force exerted on $S_V$ was strong enough to deplete clusters and decrease the presence of phase 1T which became unstable. The $S_V$ were globally displaced to the left of the clusters and an opposite voltage, -$V_H$, was required to get them to migrate in the opposite direction, corresponding to a bipolarity of the memristive effect.

We demonstrate that $S_V$ centers play a role in the slow dynamics of the memristive effect, to go further, we describe the interplay between $S_V$ defects and phase transition which is measured mechanically. Similarly to other TMDs,[29] $MoS_2$ is a polymorph 2D material, presenting a memristive crystalline phase transition[30]. The influence of some external parameters, such as doping, strain or electron injection, generate the nucleation of the metallic orthorhombic 1T phase of $MoS_2$ into the natural semiconducting 2H phase. This phase transition involves a sliding of S atoms in the unit cell. The new phase is unstable and can transit into different derivates, like for instance to the disordered 1T' phase, due to charge density wave deformations. The formation energies are 0.84 eV for (2H→1T) and 0.55eV for (2H→1T')[31]. However, to create a triangular island of 1T', this energy is insufficient: additional energy for 3 boundaries at the phase interface as well as the 3 corners is required. At the boundary between the two phases, in an S-poor $MoS_2$, there is a zig-zag chain or line of S vacancies [31], also called the α phase. The total energy cost of the nucleation in a perfect 2H-$MoS_2$ is at less 6 eV [31] which rarely occurs. However, a high density of defects, as $S_V$ and a line of $S_V$, promotes the phase transition by strongly reducing this nucleation energy and weakening the Mo-S bonds, and by acting has electron donors which inject 0.04e in the nearby Mo atom[22]. Since the S vacancies were abundant in our $MoS_2$ [32] and since they participated in the 1T' nucleation process, we intuitively assigned a high concentration of sulfur vacancy to an energy formation reduction.

Figure S9 interprets the bipolarity of the memristive effect by arguing that the migration-drift of the $S_V$ takes place in the preferential direction of the applied electric field, either in the writing phase, with the injection or the accumulation of the $S_V$ centers in α phase, or in the erasing phase, with the $S_V$ extraction of these clusters. Consequently, the $S_V$ population on either side of the cluster is modulated with the polarization. The PPC is also in line with the slow drift of $S_V$ centers acting as deep traps; their interaction with a depletion region as well as exciton dissociation and carrier collection, is favored if the $S_V$ can move over a few μm, even if this movement is extremely slow[33], until it interacts with a metallic contact, for example, and is accelerated by an applied voltage.